\documentclass[universe,review,submit,oneauthor,pdftex]{Definitions/mdpi} 

\renewcommand{\linenumbers}{}

\firstpage{1} 
\pubvolume{1}
\issuenum{1}
\articlenumber{0}
\pubyear{2021}
\copyrightyear{2021}

\pdfoutput=1

\usepackage{amsmath}

\def\laq{\raise 0.4ex\hbox{$<$}\kern -0.8em\lower 0.62ex\hbox{$\sim$}}
\def\gaq{\raise 0.4ex\hbox{$>$}\kern -0.7em\lower 0.62ex\hbox{$\sim$}}

\def\beq{\begin{equation}}
\def\eeq{\end{equation}}
\def\bea{\begin{eqnarray}}
\def\eea{\end{eqnarray}}

\def \pa {\partial}
\def \ra {\rightarrow}

\def \fb {\overline \phi}
\def \fbp {\dot{\fb}}
\def \bp {\dot{\beta}}

\def \la {\lambda}
\def \ls {\lambda_s}
\def \lp {\lambda_P}
\def \La {\Lambda}

\def \b {\beta}
\def \a {\alpha}
\def \ap {\alpha^{\prime}}

\def \ga {\gamma}
\def \sg {\sigma}
\def \da {\delta}
\def \ep {\epsilon}

\def \Om {\Omega}

\def \pfb {\Pi_{\fb}}
\def \pM {\Pi_{M}}

\Title{Quantum String Cosmology}

\Author{Maurizio Gasperini $^{1,2}$
}

\address{$^{1}$ \quad Dipartimento di Fisica, Universit\`a di Bari, 
Via G. Amendola 173, 70126 Bari, Italy\\
$^{2}$ \quad Istituto Nazionale di Fisica Nucleare, Sezione di Bari, Italy}

\abstract{We present a short review of possible applications of the Wheeler-De Witt equation to cosmological models based on the low-energy string effective action, {and characterised by an initial regime of asymptotically flat, low energy, weak coupling evolution.} 
Considering in particular a class of duality-related (but classically disconnected) background solutions, we shall discuss the possibility of quantum transitions between the phases of pre-big bang and post-big bang evolution. We will show that it is possible, in such a context, to represent the birth of our Universe as a quantum process of tunneling or ``anti-tunneling'' from an initial state asymptotically approaching the string perturbative vacuum. }

\keyword{string cosmology; quantum cosmology; Wheeler-De Witt equation.
\\ \\ 
{\bf To appear in} the Special Issue on {\em ``Quantum Cosmology"}, edited by Paulo Vargas Moniz \\ ({\em Universe}, MDPI, 2021).
\\ \\ 
{\bf Preprint}: BA-TH/802-20  
\\ \\ 
{\bf Availabe at}: https://doi.org/10.3390/universe7010014}

\begin{document}

\newpage


\section{Introduction}
\label{sec1}

In the standard cosmological context the Universe is expected to emerge from the big bang singularity and to evolve initially through a phase of very high curvature and density, well inside the quantum gravity regime. Quantum cosmology, in that context, turns out to be a quite appropriate formalism to describe the ``birth of our Universe'', possibly in a state approaching the de Sitter geometric configuration  typical of inflation (see, e.g., \cite{1} for a~review).

In the context of  string cosmology, in contrast, there are scenarios where the Universe emerges from a state satisfying the postulate of ``asymptotic past triviality'' \cite{2} (see \cite{3} for a recent discussion):  in that case the initial phase is classical, with a curvature and a density very small in string (or Planck) units. 
Even in that case, however, the transition to the decelerated radiation-dominated evolution, typical of standard cosmology, is expected to occur after crossing a regime of very high-curvature and strong coupling. The birth of our Universe, regarded as the beginning of the standard cosmological state, corresponds in that case to the transition (or ``bounce'') from the phase of growing to decreasing curvature, and even in that case can be described by using quantum cosmology methods, like for a Universe  emerging from an initial singularity.

There is, however, a crucial difference between a quantum description of the ``big bang'' and of the ``big bounce'': indeed, the bounce is preceded by a long period of low-energy, classical evolution, while the standard big bang picture implies that the space-time dynamics suddenly ends at the singularity, with no classical description  at previous epochs (actually, there are no ``previous'' epochs, as the time coordinate itself ends at the singularity). In { that}  context the initial state of the Universe is unknown, and has to be fixed through some ad hoc prescription: hence, different choices for the initial boundary conditions are in principle allowed \cite{4,5,6,7,8,9}, leading in general to different quantum pictures for the very early cosmological evolution. {Such an approach, based in particular on ``no-boundary'' initial conditions, has been recently applied also to the ekpyrotic scenario \cite{9a}, leading to the production of ekpyrotic instantons
\cite{9b,9c}.}
In the class of string cosmology models considered in this paper, in contrast, the initial state is uniquely determined by a fixed choice of pre-big bang (or pre-bounce) evolution (see, e.g., \cite{10,11,12,13,14}), which starts asymptotically from the string perturbative vacuum  and which, in this way, unambiguously determines the initial ``wave function'' of the Universe and the subsequent transition probabilities. 

In this paper we report the results of previous works,   based on the study of the Wheeler--De Witt (WDW) equation \cite{15,16} in the ``minisuperspace''  associated { with} a class of cosmological {backgrounds} 
 compatible with the dynamics the low-energy string effective action \cite{17,18,19,20,21,22,23}.
It is possible, in such a context, to obtain a non-vanishing transition probability between two different geometrical configurations---in particular, from a pre-big bang to a post-big bang state---even if they are classically disconnected by a space-time singularity. There is no need, to this purpose, of adding higher-order string-theory contributions (like $\ap$ and loop corrections) to the WDW equation, except those possibly encoded into an effective (non-local) dilaton potential (but see \cite{24,25} for high-curvature contributions to the WDW equation). It will be shown, also, that there are no problems of operator ordering in the WDW equation, as the ordering is automatically fixed by the duality symmetry of the effective action. 
Other possible problems---of conceptual nature and typical of the WDW approach to quantum cosmology---however,  remain, like the the validity of a probabilistic interpretation of the wave function \cite{26}, the existence and the possible meaning of a semiclassical limit \cite{27}, the unambiguous identification of a time-like coordinate in superspace (see however \cite{20}, and see \cite{20a} {for a recent discussion}).

{Let us stress that this review is dedicated in particular to string cosmology backgrounds of the pre-big bang type, and limited to a class of spatially homogeneous geometries. It should be recalled, however, that there are other important works in a quantum cosmology context which are also directly (or indirectly) related to the string effective action, and which are applied to more general classes of background geometries not necessarily characterised by spatial Abelian isometries, and not necessarily emerging from the string vacuum. 

We should mention, in particular,  the quantum cosmology results for the bosonic sector of the heterotic string with Bianchi-type IX geometry \cite{31a} and Bianchi class A geometry~\cite{31b}; solutions for the WDW wave function with quadratic and cubic curvature corrections \cite{31c} (typical of $f(R)$ models of gravity), describing a phase of conventional inflation; two-dimensional models of dilaton quantum cosmology and their supersymmetric extension~\cite{31d}; WDW equation for a class of scalar-tensor theories of gravity with a generalised form of scale-factor duality invariance \cite{31e,31f}. We think that discussing those (and related) works should deserve by itself a separate review paper.}

{ This} paper is organized as follows. In Section \ref{sec2} we present the explicit form of the WDW equation following from the low-energy string effective action, {for homogeneous backgrounds with $d$ Abelian spatial isometries}, and show that it is free from operator-ordering ambiguities thanks to its intrinsic $O(d,d)$ symmetry. In Section \ref{sec3}, working in the simple two-dimensional minisuperspace associated with a class of exact gravi-dilaton solutions of the string cosmology equations, we discuss the scattering of the WDW wave function induced by the presence of a generic dilaton potential. In Section \ref{sec4} we show that an appropriate quantum reflection of the wave function can be physically interpreted as representing the birth of our Universe as a process of tunnelling from the string perturbative vacuum. Similarly, in Section \ref{sec5}, we show that the parametric amplification of the WDW wave function can describe the birth of {{our}} 
 Universe as a process of ``anti-tunnelling'' from the string perturbative vacuum. Section \ref{sec6} is finally devoted to a few conclusive remarks.
 

\renewcommand{\theequation}{2.\arabic{equation}}
\setcounter{equation}{0}
\section{The Wheeler-De Witt equation for the low-energy string effective action}
\label{sec2}

In a quantum cosmology context the Universe is described by a wave function evolving in the so-called superspace and governed by the WDW equation \cite{15,16}, in much the same way as in ordinary quantum mechanics a particle is described by a wave function evolving in Hilbert space \cite{28}, governed by the Schrodinger equation. Each point of superspace corresponds to a possible geometric configuration of the space-like sections of our cosmological space-time, and the propagation of the WDW wave function through this manifold describes the quantum dynamics of the cosmological geometry (thus providing, in particular, the transition probabilities between different geometric states).

The WDW equation, which implements the Hamiltonian constraint $H=0$ in the superspace of the chosen cosmological scenario, has to be obtained, in our context, from the  appropriate string effective action. Let us consider, to this purpose, the low-energy, tree-level, $(d+1)$-dimensional (super)string effective action, which can be written as \cite{29,30,31}
\beq
S=-{1\over 2\la_s^{d-1}}\int d^{d+1}x \sqrt{-g}\,e^{-\phi}\left(R+
\pa_\mu\phi\pa^\mu\phi-{1\over 12}H_{\mu\nu\a}H^{\mu\nu\a}+V\right),
\label{21}
\eeq
where $\phi$ is the dilaton \cite{32}, $H_{\mu\nu\a}= \pa_\mu B_{\nu\a}+\pa_\nu B_{\a\mu}+\pa_\a B_{\mu\nu}$ is the field strength 
of the NS-NS two form $B_{\mu\nu}=-B_{\nu\mu}$ (also called Kalb-Ramond axion), 
and $\la_s\equiv (\ap)^{1/2}$ is the fundamental string length parameter. We have also added a possible non-trivial dilaton potential $V(\phi)$.

For the purpose of this paper it will be enough to consider a class of  homogeneous backgrounds with $d$ Abelian spatial isometries and spatial 
sections of finite volume, i.e. $(\int d^dx\sqrt{-g})_{t={\rm const}}<\infty$. In the synchronous frame where $g_{00}=1$, $g_{0i}=0=B_{0i}$, and where the fields are independent of all 
space-like coordinates $x^i$ ($i,j=1,..,d$), the above action can then be rewritten as follows \cite{33,34}:
\beq
S= \int dt \,L(\fb, M), ~~~~~~~~~~~~~
L=-{\ls\over 2}e^{-\fb}\left[(\dot{\fb})^2+{1\over 8}{\rm Tr}
~\dot M(M^{-1})\dot{}+V\right].
\label{22}
\eeq
Here a dot denotes differentiation with respect to the cosmic time $t$, 
and $\fb$ is the so-called  ``shifted" dilaton field,
\beq
\fb=\phi-\ln \sqrt{-g},
\label{23}
\eeq
where we have absorbed into $\phi$ the constant shift $-\ln(\ls^{-d}\int 
d^dx)$. Finally, $M$ is the $2d\times 2d$ matrix
\beq
M=\begin{pmatrix}
G^{-1} & -G^{-1}B \\
BG^{-1} & G-BG^{-1}B \\
\end{pmatrix},
\label{24}
\eeq
where $G$ and $B$ are, respectively, $d\times d$ matrix representations of the spatial 
part of the metric ($g_{ij}$) and of the antisymmetric tensor ($B_{ij}$). 
For constant $V$, or for $V=V(\fb$), the above action (\ref{22}) is invariant 
under global $O(d,d)$ transformations \cite{33,34} that leave the shifted dilaton invariant, and that are parametrized in general by a constant matrix $\Om$ such that 
\beq
\fb \ra \fb , ~~~~~~~~~~~~~~~~~ M\ra \Om^T M \Om, 
\label{25}
\eeq
where $\Om$ satisfies
\beq
\Om^T\eta \Om =\eta, ~~~~~~~~~~~~~~~ \eta =
\begin{pmatrix}
0 & I \\ I & 0 
\end{pmatrix},
\label{26}
\eeq
and $I$ is the $d$-dimensional identity matrix. It can be easily checked that, in the particular case in which $B=0$ and $\Om$ coincides with $\eta$, Eq. (\ref{25}) reproduces the well-known transformation of scale-factor duality symmetry \cite{35,36}.

From the effective Lagrangian (\ref{22}) we can now obtain the (dimensionless) canonical momenta
\beq
\pfb ={\da L \over \da \dot{\fb}}= -\ls \dot{\fb}e^{-\fb}  ,~~~~~~~~~~~ 
\pM={\da L \over \da \dot M}= {\ls \over 8}e^{-\fb} M^{-1}\dot M M^{-1} ,
\label{27}
\eeq
and the associated classical Hamiltonian:
\beq
H= {e^{\fb}\over 2 \ls}\left[-\pfb^2+8 {\rm Tr}\left(M\,\Pi_M \,M\,\Pi_M\right)
+{\ls^2}V e^{-2\fb}\right].
\label{28}
\eeq
The corresponding WDW equation, implementing in superspace the Hamiltonian constraint $H=0$ through the differential operator representation $\pfb =\pm i \da/\da\fb$, 
$\pM =\pm i \da /\da M$, would seem thus to be affected  by the usual
problems of operator ordering, since $[M,\pM]\not= 0$.

The problem disappears, however, if we use the $O(d,d)$ covariance of the action (\ref{22}), and the symmetry properties of the axion-graviton field represented by the matrix (\ref{24}), which satisfies the identity $M\eta= \eta M^{-1}$. 
Thanks to this property, in fact, we can identically rewrite the axion-graviton part of the kinetic term appearing in the Lagrangian (\ref{22}) as follows:
\beq
{\rm Tr}~ \dot M(M^{-1})\dot{}=
{\rm Tr}\left(\dot M\eta \dot M \eta\right).
\label {29}
\eeq
The corresponding canonical momentum becomes
\beq
\Pi_M=-{\ls\over 8}\,e^{-\fb}\, \eta \dot M \eta,
\label{210}
\eeq
and the associated Hamiltonian
\beq
H= {e^{\fb}\over 2 \ls}\left[-\pfb^2-8 {\rm Tr}\left(\eta\,\Pi_M \,\eta\,\Pi_M\right)
+{\ls^2}V e^{-2\fb}\right]
\label{211}
\eeq
has a flat metric in momentum space, and leads to a WDW equation
\beq
\left[{\da^2 \over \da \fb^2}+ 8{\rm Tr}\left(\eta {\da \over \da M}
\eta {\da \over \da M}\right) +\ls^2 V e^{-2\fb}~ \right]\Psi(\fb, M)=0
\label{212}
\eeq
which is manifestly free from problems of operator ordering.

Finally, it may be interesting to note that the quantum ordering imposed by the dualiy symmetry of the effective action is exactly equivalent to the order fixed by  the condition of reparametrization invariance in superspace. 

To check this point let us consider a simple spatially isotropic background, with $B_{ij}=0$ and scale factor $a(t)$, so that $G_{ij}=-a^2(t) \da_{ij}$. The effective Lagrangian (\ref{22}) becomes
\beq 
L(\fb, a)= -{\ls\over 2}e^{-\fb}\left(\dot{\fb}^2- d{\dot a^2\over a^2} +V \right),
\label{213}
\eeq
with associated canonical momenta
\beq
\pfb ={\da L \over \da \dot{\fb}}= -\ls \dot{\fb}e^{-\fb}  ,~~~~~~~~~~~ 
\Pi_a={\da L \over \da \dot a}= \ls   d{\dot a\over a^2}   e^{-\fb}  ,
\label{214}
\eeq
and Hamiltonian constraint:
\beq
2 \ls e^{-\fb}H= -\pfb^2+ {a^2\over d} \Pi_a^2 +{\ls^2}V e^{-2\fb}=0.
\label{215}
\eeq
The differential implementation of this constraint in terms of the operators $\pfb \ra \pm i \pa/\pa \fb$, $\Pi_a \ra \pm i \pa/\pa a$ has to be ordered, because $[a, \Pi_a]\not=0$. It follows that in general, for the kinetic part of the Hamiltonian $H_k= -\pfb^2+ a^2 \Pi_a^2/d$, we have the following differential representation
\beq
H_k= {\pa\over \pa \fb^2}-{a^2\over d} {\pa^2 \over \pa a^2} - \ep {a\over d} {\pa \over \pa a},
\label{216}
\eeq
where $\ep$ is a numerical parameter depending on the imposed ordering. However, if we perform a scale-factor duality transformation  $\fb \ra \fb$,  $a \ra \widetilde a =a^{-1}$ (which exactly corresponds to the class of transformations (\ref{25}) for the particular class of backgrounds that we are considering), we find
\beq
H_k(a)= H_k(\widetilde a) +{2\over d} (\ep-1)\, \widetilde a \,{\pa\over \pa \widetilde a}.
\label{217}
\eeq
{The duality} invariance of the Hamiltonian thus requires $\ep=1$ (which, by the way, is also the value of $\ep$ that we have to insert into Equation (\ref{216}) to be in agreement with the general result (\ref{212}) if we consider the particular class of geometries with $B=0$ and $G=-a^2 I$). {See also \cite{31e,31f} for the WDW equation with a generalised form of scale-factor duality~symmetry.}

Let us now consider the kinetic part of the Hamiltonian operator (\ref{215}), which is given as a quadratic form in the canonical momenta $\Pi_A= (\pfb, \Pi_a)$, written in a 2-dimensional minisuperspace with a non-trivial metric $\ga_{AB}$ and coordinates $x^A= (\fb, a)$, such that:
\beq
H_k= - \Pi_{\fb}^2+{a^2\over d} \Pi_a^2 \equiv \ga^{AB} \Pi_A\Pi_B,
~~~~~~~~~~~
\ga_{AB}(\fb, a)= {\rm diag} \left(-1,{d\over a^2}\right).
\label{218}
\eeq
If we impose on the differential representation of the Hamiltonian constraint the condition of general covariance with respect to the given minisuperspace geometry \cite{37}, 
we  obtain
\beq 
H_k =-\ga^{AB} \nabla_A \nabla_B =
-{1\over \sqrt{-\ga}} \pa_A (\sqrt{-\ga} \ga^{AB} \pa_B)  \equiv
{\pa\over \pa \fb^2}-{a^2\over d} {\pa^2 \over \pa a^2} -  {a\over d} {\pa \over \pa a},
\label{219}
\eeq
and this result exactly reproduces the differential operator (\ref{216}) with $\ep=1$. The duality symmetry of the action, and the requirement of reparametrisation invariance in superspace,  are thus equivalent to select just the same ordering prescription, as previously anticipated.


\renewcommand{\theequation}{3.\arabic{equation}}
\setcounter{equation}{0}
\section{Quantum scattering of the Wheeler-De Witt wave function in minisuperspace}
\label{sec3}

For an elementary discussion of this topic, and for the particular applications we have in mind -- namely, a quantum description of the ``birth" of our present cosmological state from the string perturbative vacuum -- we shall consider the homogeneous, isotropic and spatially flat class of $(d+1)$-dimensional backgrounds already introduced in the previous section, with $B_{\mu\nu}=0$, $g_{00}=1$ and scale factor $a(t)$. We shall thus work in a  two-dimensional minisuperspace, spanned by the convenient coordinates 
$(\fb, \b)$ where $\b=\sqrt d \ln a$. With such variables the effective Lagrangian (\ref{22})  takes the form
\beq
L(\b,\fb)=-\ls\,{e^{-\fb}\over 2}\,
\left[\dot{\fb}^2 - \dot{\beta}^2+
V (\b, \fb)\right], 
\label{31}
\eeq
and the momenta, canonically conjugate to the coordinates $\fb$, $\b$, are given by
\beq
\pfb ={\da L \over \da \dot{\fb}}= -\ls \dot{\fb}e^{-\fb}  ,~~~~~~~~~~~~~~~~~ 
\Pi_\b={\da L \over \da \dot \b}=\lambda_{\rm s}\,\dot{\beta}\,e^{-\fb}.
\label{32}
\eeq
The Hamiltonian constraint (\ref{215}) becomes 
\beq
-\Pi^2_{\fb}+\Pi^2_{\beta}
+\lambda_{\rm s}^2\,V(\b,\fb)\,e^{-2\,\fb}=0, 
\label{33}
\eeq
corresponding to an effective WDW equation
\beq
\left[ \partial^2_{\fb} - 
\partial^2_{ \beta}
+\lambda_{\rm s}^2\,V(\b,\fb)\,e^{-2\fb}~ \right ] \Psi(\b,\fb)= 0 . 
\label{34}
\eeq
For $V=0$ we have the free D'Alembert equation, and the general solution can be written in terms of plane waves as
\beq
\Psi(\b,\fb)= \psi^\pm_\b \psi^\pm_{\fb} \sim e^{\mp i k \b} e^{\mp i k \fb}.
\label{35}
\eeq
Here $k>0$, and $\psi^\pm_{\b}$, $\psi^\pm_{\fb}$ are free momentum eigenstates, satisfying the eigenvalue equations
\beq
\Pi_{\b}\,\psi_{\b}^{\pm}= \pm k \,\psi_{\b}^{\pm} , 
~~~~~~~~~~~~~~~
\Pi_{\fb}\,\psi_{\fb}^{\pm}= \pm k \,\psi_{\fb}^{\pm}.
\label{36}
\eeq

Let us now recall that, for $V=0$, the equations following from the effective Lagrangian (\ref{31}) admit a class of exact solution describing four (physically different) cosmological phases, two expanding and two contracting, parametrized by \cite{10,12,13,14}:
\beq
a(t)\sim (\mp t)^{\mp1/\sqrt d} , ~~~~~~~~~~~~~~
\fb (t) \sim  -\ln (\mp t) , 
\label{37}
\eeq
They are defined on the disconnected time ranges $[- \infty, 0[$ and $]0, +\infty]$, and are related by duality transformations $a \ra a^{-1}$, $\fb \ra \fb$ and time-reversal transformation, $t \ra -t$. They may represent the four asymptotic branches of the low-energy string cosmology solutions even in the presence of a non-vanishing dilaton potential, provided the effective contribution of the potential is localized in  a region of finite extension of the $(\fb, \b)$ plane, and goes (rapidly enough) to zero as $\fb, \b \ra \pm \infty$. 

The above solutions satisfy the condition
\beq
\fbp= \pm \sqrt d \,{\dot a \over a} = \pm \bp ,
\label{38}
\eeq
so that, according to the definitions (\ref{32}), they correspond to configurations with canonical momenta related by $\Pi_\b = \pm \Pi_{\fb}$. By recalling that the phase of (expanding or contracting) pre-big bang evolution is characterized by growing curvature and growing dilaton  \cite{12,13} (namely, $\dot \fb >0$, $\Pi_{\fb}<0$), while the curvature and the dilaton are decreasing in the (expanding or contracting) post-big bang phase (where $\dot \fb <0$, $\Pi_{\fb}>0$), we can conclude, according to Eqs. (\ref{32}), (\ref{36}), that the classical solutions (\ref{37}) of the string cosmology equations admit the following plane-wave representation in minisuperspace in terms of $\psi^\pm_{\b}$, $\psi^\pm_{\fb}$:
\begin{itemize}
\item{}expansion ~~~~~~$ \longrightarrow ~~~\bp >0~~~ \longrightarrow ~~~\psi_\b^+$,
\item{}contraction~~~~ $\longrightarrow ~~~\bp <0 ~~~\longrightarrow ~~~\psi_\b^-$,
\item{}pre-big bang (growing dilaton) ~~~~~~~ $\longrightarrow ~~~ \fbp >0~~~\longrightarrow ~~~\psi_{\fb}^-$,
\item{}post-big bang (decreasing dilaton) ~~$\longrightarrow ~~~ \fbp <0~~~\longrightarrow~~~ \psi_{\fb}^+$. 
\end{itemize}

Let us now impose, as our physical boundary condition,  that the initial state of our Universe describes a phase of expanding pre-big bang evolution, asymptotically
emerging from the string perturbative vacuum (identified with the limit $\b \ra
-\infty$,  $\phi \ra -\infty$). It follows that the initial state $\Psi_{\rm in}$ must
represent a configuration with $\dot \b>0$ and $\dot \fb >0$, namely a state with 
positive eigenvalue of $\Pi_\b$ and negative 
(opposite) eigenvalue of $\Pi_{\fb}$, i.e. $\Psi_{\rm in} \sim \psi_{\b}^{+}
\psi_{\fb}^{-}$. 

In such a context, a quantum  transition from the pre- to the post-big bang regime can be described as a  process of scattering of the initial wave function induced by the presence of some appropriate dilaton potential, which we shall assume to have non-negligible dynamical effects only in a finite region localized around the origin of the minusuperspace spanned by the $(\fb, \b)$ coordinates. In other words, we shall assume that the contributions of $V(\phi)$ to the WDW equation tend to disappear not only in the initial but also in the final asymptotic regime where $\b \ra
+ \infty$,  $\phi \ra +\infty$. As a consequence, also the final asymptotic configuration $\Psi_{\rm out}$, emerging from the scattering process, can be represented in terms of the free momentum eigenstates $\psi^\pm_{\b}$ and $\psi^\pm_{\fb}$. 

\begin{figure}[h]
\centering
\includegraphics[width=12 cm]{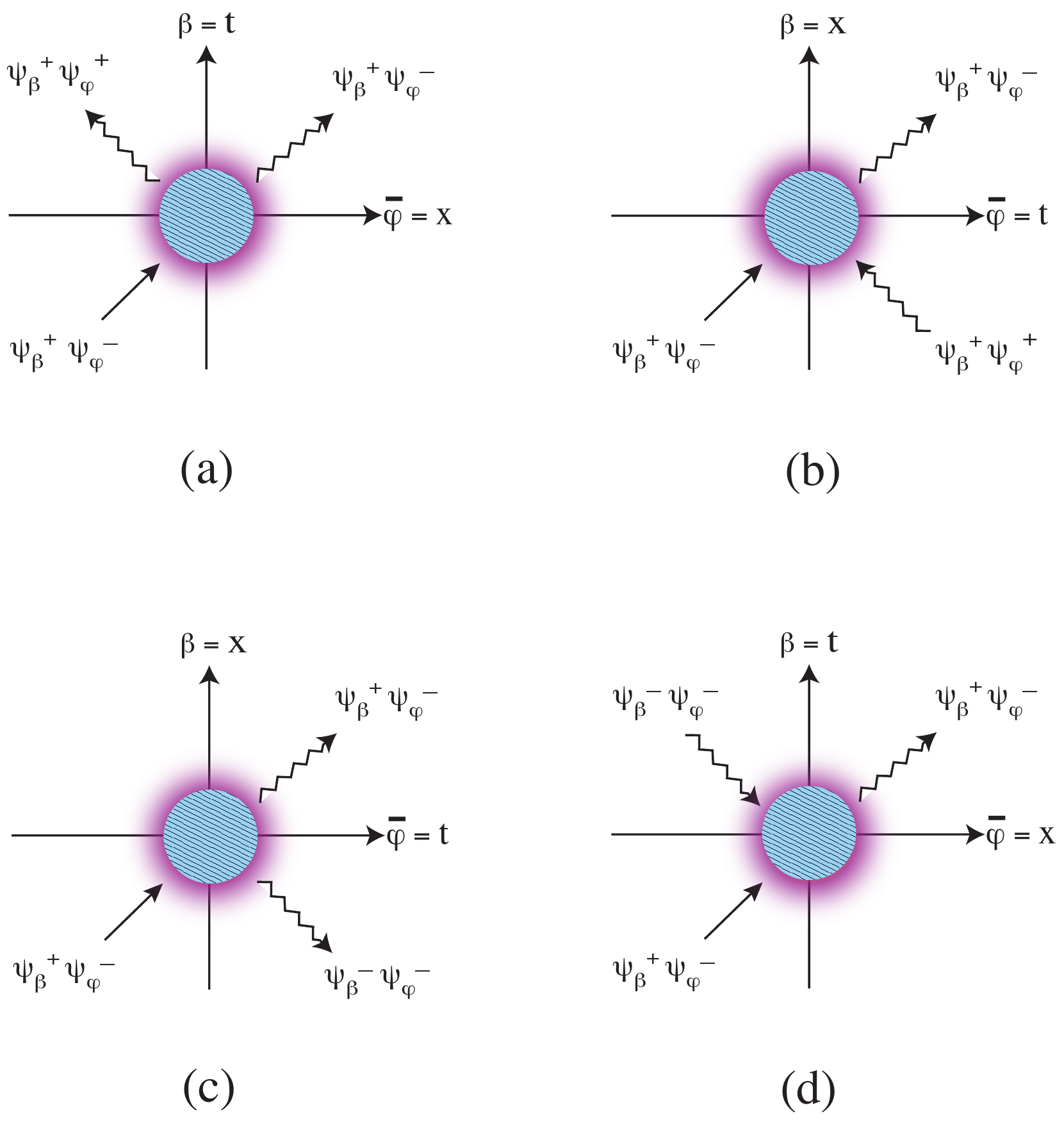}
\caption{Four different classes of scattering processes for the
incoming wave function describing a phase of expanding pre-big bang evolution, asymptotically
emerging from the string perturbative vacuum
(straight, solid line). The outgoing state is represented by a mixture of
eigenfunctions of $\Pi_\b$ and $\Pi_{\fb}$ with positive and negative eigenvalues.  See the main text for a detailed explanation of the four different cases (a)--(d) illustrated in this figure.}
\label{f1}
\end{figure}

However, unlike the initial state fixed by the chosen boundary conditions -- and selected to represent a configuration with $\Pi_\b >0$ and $\Pi_{\fb}<0$ -- the final state is not constrained by such a restriction and can describe in general different  configurations. In particular, the scattering process may lead to configurations asymptotically described by a wave function $\Psi_{\rm out}$ which is a superposition of different momentum eigenstates: for instance, waves with the same 
$\Pi_\b >0$ and opposite values of $\Pi_{\fb}$, i.e. $\Psi_{\rm out}\sim \psi^+_{\b}\psi^\pm_{\fb}$ (see Fig. \ref{f1}, cases (a) and (b)); or waves  with the same 
$\Pi_{\fb}< 0$ and opposite values of $\Pi_{\b}$, i.e. $\Psi_{\rm out}\sim \psi^-_{\fb}\psi^{\pm}_{\b}$ (see Fig. \ref{f1}, cases (c) and (d)).

It may be interesting to note that those different configurations may be interpreted as different possible ``decay channels" of the string perturbative vacuum \cite{21}. Also, it should be stressed  (as clearly illustrated in Fig. \ref{f1}) that one of the two components of the outgoing wave function $\Psi_{\rm out}$ must always correspond to the ``transmitted" part of the incident wave $\Psi_{\rm in}$, namely must correspond to a state with $\Pi_\b>0$ and $\Pi_{\fb}<0$, represented by $ \psi_{\b}^{+}\psi_{\fb}^{-}$. However, the ``reflected" part of the wave function may have different physical interpretations, also depending on the chosen identification of the time-like coordinate in minisuperspace \cite{20,22}: the $\b$ axis for the cases (a) and (d), the $\fb$ axis for the case (b) and (c).

It turns out that only the cases (a)  and (c) of Fig. \ref{f1} represent a true process of reflection of the incident wave along
a spacelike coordinate (the axes $\fb$ and $\b$, respectively).  In case (a), in particular, the
evolution along $\b$ is monotonic, the Universe always keeps 
expanding, and the incident wave $\Psi_{\rm in}$  is partially transmitted towards the
pre-big bang singularity (with unbounded growth of the curvature and of
the dilaton, $\b \ra +\infty$, $\fb \ra +\infty$), and partially reflected
back towards the expanding, low-energy, post-big bang regime  ($\b
\ra +\infty$, $\fb \ra -\infty$). As we shall show in Sect. \ref{sec4}, this type of quantum reflection can also be interpreted as a process of ``tunnelling" from the string perturbative vacuum. 

The cases (b) and (d) of Fig. \ref{f1} are qualitatively different, as the
final state  is a superposition of modes of positive and negative frequency with respect to the chosen timelike coordinate (the axes $\fb$ and $\b$, respectively). Namely, $\Psi_{\rm out}$ is a superposition of positive and negative energy
eigenstates, and this represents a quantum process of ``parametric amplification" of the wave function \cite{38,39} or, in the language of third quantization
\cite{40,41,42,43,44}  -- i.e. second quantization of
the WDW wave function -- a process of  ``Bogoliubov
mixing" of the energy modes (see e.g. \cite{45,46}), associated with the production of ``pairs of universes" from the vacuum. For that process, the mode ``moving backwards" with respect to the chosen time coordinate has to be ``reinterpreted": as an anti-particle in the usual quantum field theory context,  as an ``anti-universe" in a quantum cosmology context. 

Such a re-interpretation principle produces, as usual, states of
positive energy and opposite momentum. It turns out, in particular, that the case (d) of Fig. \ref{f1} describes -- after the correct re-interpretation -- the production of  universe/anti-universe pairs  in which both members of the pair have positive energy and positive momentum along the $\b$ axis. Hence, they are both expanding:  one falls
inside the pre-big bang singularity ($\fb \ra +\infty$), but the other expands towards the
low-energy post-big bang regime ($\fb \ra -\infty$). As we shall show in Sect. \ref{sec5}, this quantum effect of pair production can also be interpreted as a process of ``anti-tunnelling" from the string perturbative vacuum.


\renewcommand{\theequation}{4.\arabic{equation}}
\setcounter{equation}{0}
\section{Birth of the Universe as a tunnelling from the string perturbative vacuum}
\label{sec4}

To illustrate the process of quantum transition from the pre- to the post-big bang regime as a wave reflection in superspace we shall consider here the simplest (almost trivial) case of constant dilaton potential, $V=V_0=$ const (see e.g. \cite{17} for more general dynamical configurations). With this potential the classical background solutions for the cosmological equations of the effective Lagrangian (\ref{31}) are well known \cite{47}, and can be written as
\beq
a(t)= a_0 \left[\tanh\left(\mp \sqrt{V_0} t/2\right) \right]^{\mp 1/\sqrt d},
~~~~~~~~
\fb=\phi_0- \ln \left[\sinh \left(\mp \sqrt{V_0} t \right)\right],
\label{41}
\eeq
where $a_0$ and $\phi_0$ are integration constants. 

These solutions have two branches, of the pre-big bang type ($\dot \fb >0$) and post-big bang type ($\dot \fb <0$), defined respectively over the disconnected time ranges $t<0$ and $t>0$, and classically separated by a singularity of the curvature  and of the effective string coupling ($\exp \fb$)  at $t=0$. For $t \ra \pm \infty$ they approach, asymptotically, the free vacuum solution (\ref{37}) obtained for $V=0$. It is important to note, also, that each branch of the above solution can describe either expanding or contracting geometric configurations, which are both characterized by a constant canonical momentum along the $\b$ axis, given (according to Eq. (\ref{32})), by 
\beq
\Pi_\b= \ls \dot{\b}\,e^{-\fb}  = \pm k, ~~~~~~~~~~~~
k=  \ls \sqrt{V_0} \,e^{-\phi_0}.
\label{42}
\eeq

Let us now apply the WDW equation (\ref{34}) to compute the (classically forbidden) probability of transition  from the pre- to the post-big bang branches of the solution (\ref{41}). We are interested, in particular, in the transition between expanding configurations, and we shall thus consider the quantum process described by the case $(a)$ of Fig. \ref{f1}, with a wave function monotonically evolving along the positive direction of the $\b$ axis (also in agreement with the role of time-like coordinate asssigned to $\b$). In that case $\dot \b>0$, and the conserved canonical momentum (\ref{42}) is positive, $\Pi_\b>0$. By imposing momentum conservation as a differential condition on the wave function,
\beq
\Pi_\b \Psi_k (\b, \fb) = i \pa_\b  \Psi_k (\b, \fb) = k\, \Psi_k (\b, \fb),
\label{43}
\eeq
we can then separate the variables in the solution of the WDW equation (\ref{34}), and we obtain
\beq
\Psi(\b, \fb)=e^{-ik \b} \psi_k(\fb) ,~~~~~~~~~~~~
\left(\pa^2_{\fb}+k^2 + \ls^2V_0 e^{-2 \fb} \right) \psi_k( \fb)=0.
\label{44}
\eeq

The general solution of the above equation can now be written as a linear combination of Bessel functions \cite{48}, $A J_\nu(z)+ BJ_{-\nu}(z)$, of index $\nu= ik$ and argument $z= \ls \sqrt{V_0} \exp(-\fb)$. Consistently with the chosen boundary conditions for the process illustrated in case $(a)$ of Fig. \ref{f1} (namely, with the choice of an initial wave function asymptotically incoming from the string perturbative vacuum), we have now to impose that there are only right-moving waves (along $\fb$) approaching the high-energy region and the final singularity in the limit $\b \ra +\infty$, $\fb \ra + \infty$. Namely, waves of the type $\psi_{\fb}^-$ -- see Eqs. (\ref{35}) and (\ref{36}) -- representing a state with $\fbp>0$ and $\Pi_{\fb}<0$. By using the small  argument limit of the Bessel functions \cite{48}, 
\beq
\lim_{\fb \ra + \infty} J_{\pm ik} \left(\ls \sqrt{V_0}\, e^{-\fb}\right) \sim e^{\mp i k \fb},
\label{45}
\eeq
we can then eliminate the   $J_\nu (z)$ component and uniquely fix the WDW solution (modulo an arbitrary normalization factor $N_k$) as follows:
\beq
\Psi_k(\b, \fb)=N_k J_{-ik}  \left(\ls \sqrt{V_0}\, e^{-\fb}\right)e^{-ik\b}.
\label{46}
\eeq

Let us now consider the wave content of this solution in the opposite, low-energy limit $\fb \ra -\infty$, where the large argument limit of the Bessel functions gives \cite{48}
\beq
\lim_{\fb \ra - \infty} \Psi_k(\b, \fb)={N_k\, e^{-ik \b}\over (2 \pi z)^{1/2} }\left[e^{-i(z-\pi/4)}e^{k\pi/2} +e^{i(z-\pi/4)}e^{-k\pi/2} \right]
\equiv \Psi_k^-(\b, \fb)+\Psi_k^+(\b, \fb),
\label{47}
\eeq
and where the two wave components $\Psi_k^-$ and $ \Psi_k^+$ are asymptotically eigenstates of $\Pi_{\fb}$ with negative and positive eigenvalues, respectively. Hence, we find in this limit a superposition of right-moving and left-moving modes (along $\fb$), representing, respectively, the initial, pre-big bang incoming state $\Psi_k^-$ (with $\Pi_{\fb}<0$, i.e. growing dilaton), and the final, post-big bang reflected component $\Psi_k^+$ (with $\Pi_{\fb}>0$, i.e. decreasing dilaton). Starting from an initial pre-big bang configuration, we can then obtain a finite probability for the transition to the ``dual" post-big bang regime, represented as a reflection of the wave function in minisuperspace, with reflection coefficient
\beq
R_k= {\left|\Psi_k^+(\b, \fb)\right|^2\over \left|\Psi_k^-(\b, \fb)\right|^2}
= e^{-2\pi k}.
\label{48}
\eeq
The  probability for this quantum process is in general nonzero, even if the corresponding  transition is classically forbidden.  

It may be interesting to evaluate $R_k$ in terms of string-scale variables, for a  region of $d$-dimensional space of given proper volume $\Om_s$. By computing the constant momentum $k$ of Eq. (\ref{42}) at the string epoch $t_s$, when $\dot \b (t_s)= \sqrt{d} (\dot a /a)(t_s) \simeq  \sqrt{d} \la_s^{-1}$, and using the definition (\ref{23}) of $\fb$, we find
\beq
R_k \sim \exp\left\{-{2 \pi \sqrt{d}\over g_s^2}{\Om_s\over \la_s^d}\right\},
\label{49}
\eeq
where the proper spatial volume is given by $\Om_s= a^d(t_s) \int d^d x$, and where $g_s = \exp (\phi_s/2)$ is the effective value of the string coupling when the dilaton has the value $\phi_s \equiv \phi(t_s)$. Note that, for values of the coupling $g_s \sim 1$, the above probability is of order one for the formation of spacelike ``bubbles" of unit size (or smaller) in string units. In general, the probability has a typical ``instanton-like" dependence on the coupling constant, $R_k \sim \exp (g_s^{-2})$.

It may be observed, finally, that an exponential dependence of the transition probability is also typical of tunnelling processes (induced by the presence of a cosmological constant $\La$) occurring in the context of standard quantum cosmology, where the tunnelling probability can be estimated as \cite{6,26,49,50}
\beq
P \sim  \exp\left\{- {4\over \lp^2 \La}\right \}
\label{410}
\eeq
($\lp$ is the Planck length). That scenario is different because, in that case, the Universe emerges from the quantum era in a classical inflationary configuration, while, in the string scenario, the Universe is expected {\em to exit} (and not to enter) the phase of inflation thanks to quantum cosmology  effects. 

In spite of such important differences, it turns out that the string cosmology result (\ref{49}) is formally very similar to the result concerning the probability that  the birth of our Universe may be described as a quantum process of ``tunneling from nothing"  \cite{6,26,49,50}. The explanation of this formal coincidence is simple, and based on the fact that
the choice of the string perturbative vacuum as initial boundary condition implies -- as previously stressed -- that the are  only outgoing (right-moving) waves approaching the singularity at $\fb \ra +\infty$. This is exactly equivalent to imposing tunneling boundary conditions, that select {\em ``... only outgoing modes at the singular space-time boundary"} \cite{26,50}. In this sense, the process illustrated in this Section can also be interpreted as a tunneling process, not ``from nothing" but ``from the string perturbative vacuum".


\section{Birth of the Universe as anti-tunnelling from the string perturbative vacuum}
\label{sec5}
\renewcommand{\theequation}{5.\arabic{equation}}
\setcounter{equation}{0}

In this Section we shall illustrate the possible transition from the pre- to the post-big bang regime as a process of parametric amplification of the WDW wave function, also equivalent -- as previously stressed -- to a quantum process of pair production from the vacuum. We shall consider, in particular, an example in which both the initial and final configurations are expanding, like in the case $(d)$ of Fig. \ref{f1}.

What we need, to this purpose, is a WDW equation with the appropriate dilaton potential, able to produce an outgoing configuration which is a superposition of states with positive and negative eigenvalues of the momentum $\Pi_\b$ (see Fig. \ref{f1}). Since we are starting from an initial expanding (pre-big bang) regime, it follows that the contribution of the potential has to break the translational invariance of the effective Hamiltonian (\ref{33}) along the $\b$ axis (i.e. $[\Pi_\b, H] \not=0$), otherwise the final configuration described in case $(d)$ of Fig. \ref{f1} would be forbidden by momentum conservation. 

We shall work here with the simple two-loop dilaton potential already introduced in \cite{23}, possibly induced by an effective cosmological constant $\La>0$,  appropriately suppressed in the low-energy, classical regime, and given explicitly by
\beq
V(\b, \fb)= \Lambda\,\theta(-\b) \,e^{2\phi}= 
\Lambda\,\theta(-\b) \,e^{2\fb+ 2 \sqrt{d} \b}.
\label{51}
\eeq
The Heaviside step function $\theta$ has been inserted to mimic an efficient damping of the potential outside the interaction region (in particular, in the large radius limit $\b \ra +\infty$ of the expanding post-bb configuration). The explicit form and mechanism of the damping, however, is not at all a crucial ingredient of our discussion, and other, different forms of damping would be equally appropriate. 

With the given potential (\ref{51}) the effective Hamiltonian is no longer translational invariant along the $\b$ direction, but we still have $[\Pi_{\fb}, H]=0$, so that we can conveniently separate the variables in the WDW equation (\ref{34}) by factorizing the eigenstates (\ref{36}) of the canonical momentum $\Pi_{\fb}$, and we are lead to
\beq
\Psi(\b,\fb)= e^{ik\fb} \psi_k(\b), ~~~~~~~~~~
\left [\partial^2_{\b}
+k^2-\lambda_s^2\,\Lambda\,\theta(-\b)\,e^{2\sqrt{d}\b} \right]
\psi_k(\b)= 0.
\label{52}
\eeq
The above WDW equation can now be exactly solved by separately considering the two ranges of the ``temporal coordinate" $\b$, namely $\b<0$ and $\b>0$.

For $\b<0$ the contribution of the potential is non vanishing, and the general solution is a linear combination of Bessel functions $J_\mu(\sg)$ and $J_{-\mu}(\sg)$, of index  $\mu=  ik/ \sqrt{d}$ and argument $\sg= i\la_s \sqrt{\La/d}~e^{\sqrt{d}\b}$.
As before, we have to impose our initial boundary conditions requiring that, in the limit $\b \ra -\infty$, the solution may asymptotically represent a low-energy pre-big bang configuration with $\Pi_\b= -\Pi_{\fb}=k$, namely (according to Eqs. (\ref{35}), (\ref{36})):
\beq
 \lim_{\b \ra -\infty} \Psi(\b,\fb)
 \sim \psi_\b^+ \psi_{\fb}^- \sim e^{ik\fb - ik\b} .
\label{53}
\eeq
By using the small argument limit (\ref{45}), and imposing the above condition, we can then uniquely fix (modulo a normalization factor $N_k$) the solution of the WDW equation (\ref{52}), for $\b<0$, as follows: 
\beq
\Psi_k(\b,\fb)=e^{ ik\fb}\,N_kJ_{-{ik/\sqrt{d}}}\left(i\la_s
\sqrt{\La/d}~e^{\sqrt{d}\b}\right),
~~~~~~~~~~~~~~\b<0.
\label{54}
\eeq

In the complementary regime $\b>0$ the potential (\ref{51}) is exactly vanishing, and the general outgoing solution is a linear superposition of eigenstates of $\Pi_\b$ with positive and negative eigenvalues, represented by the frequency modes $\psi_\b^{\pm}$ of Eqs. (\ref{35}), (\ref{36}). We can then write
\beq
\Psi_k(\b,\fb)= e^{ ik\fb}
\left[A_+(k) e^{ -ik\b}+A_-(k) e^{ik\b}\right],
~~~~~~~~~~~~~~~\b>0 ,
\label{11}
\eeq
and the numerical coefficients $A_{\pm}(k)$ can be fixed by the two matching conditions imposing the continuity of $\Psi_k$ and $\pa_\b \Psi_k$ at $\b=0$. 

Let us now recall that the so-called Bogoliubov coefficients $|c_\pm(k)|^2 = 
|A_\pm(k)|^2/ |N_k|^2$, determining the mixing of positive and negative energy modes in the asymptotic outgoing solution \cite{45,46}, also play the role of destruction and creation operators in the context of the third quantization
formalism \cite{40,41,42,43,44}, thus controlling the ``number of universes" $n_k= |c_-(k)|^2$  produced from the vacuum, 
for each mode $k$. It turns out, in particular, that such a transition from the initial vacuum to the final standard regime, represented as a quantum scattering of the initial wave function, is an efficient process only when the final wave function is not damped but, on the contrary, turns out to be ``parametrically amplified" by the interaction with the effective potential barrier \cite{38,39}. This is indeed what happens for the solutions of our WDW equation (\ref{52}), provided the dilaton potential satisfies the condition $k < \la_s \sqrt{\La}$ \cite{23} (as can be checked by an explicit computation of our coefficients $A_\pm(k)$). 

\begin{figure}[h]
\centering
\includegraphics[width=6 cm]{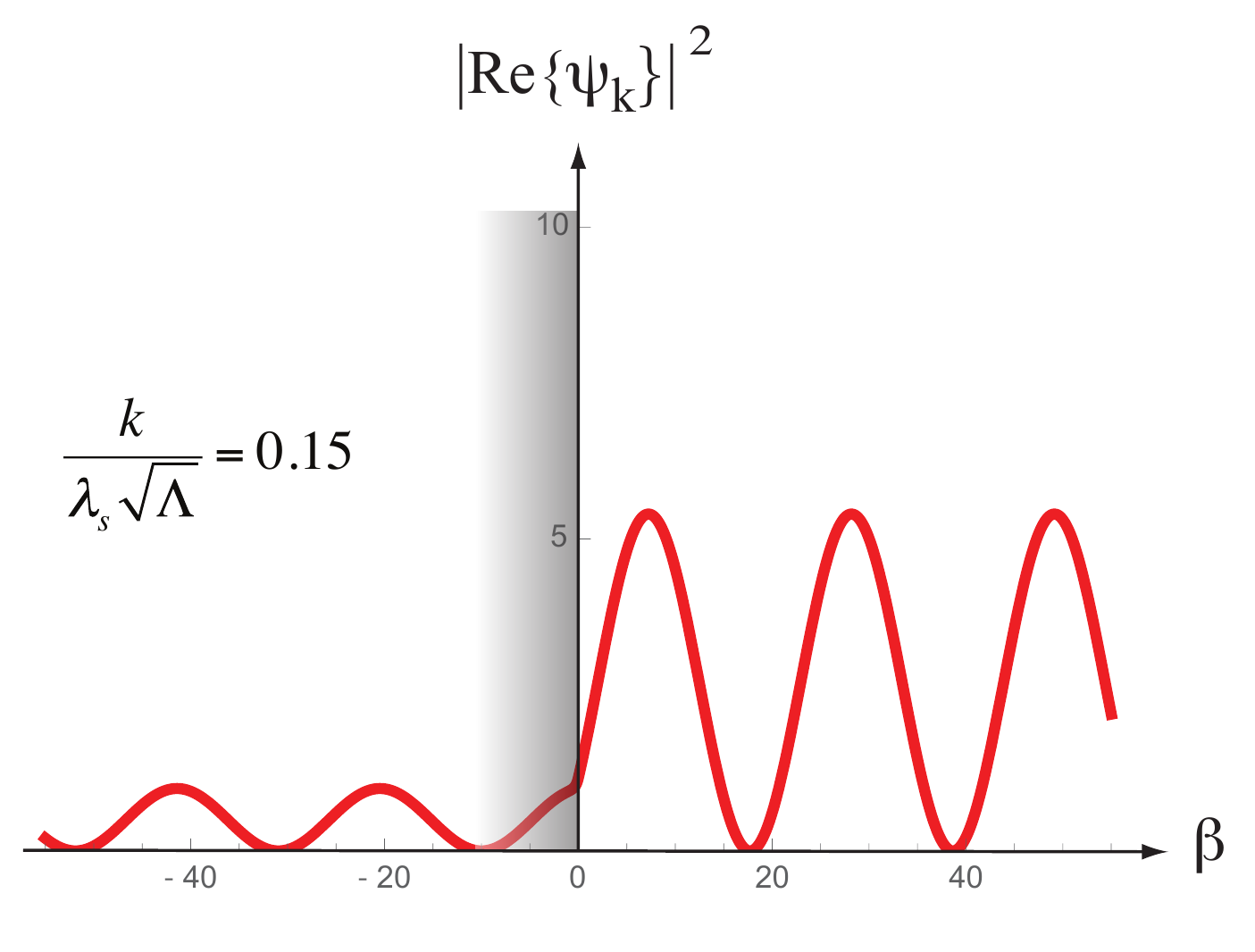}~~~
\includegraphics[width=6 cm]{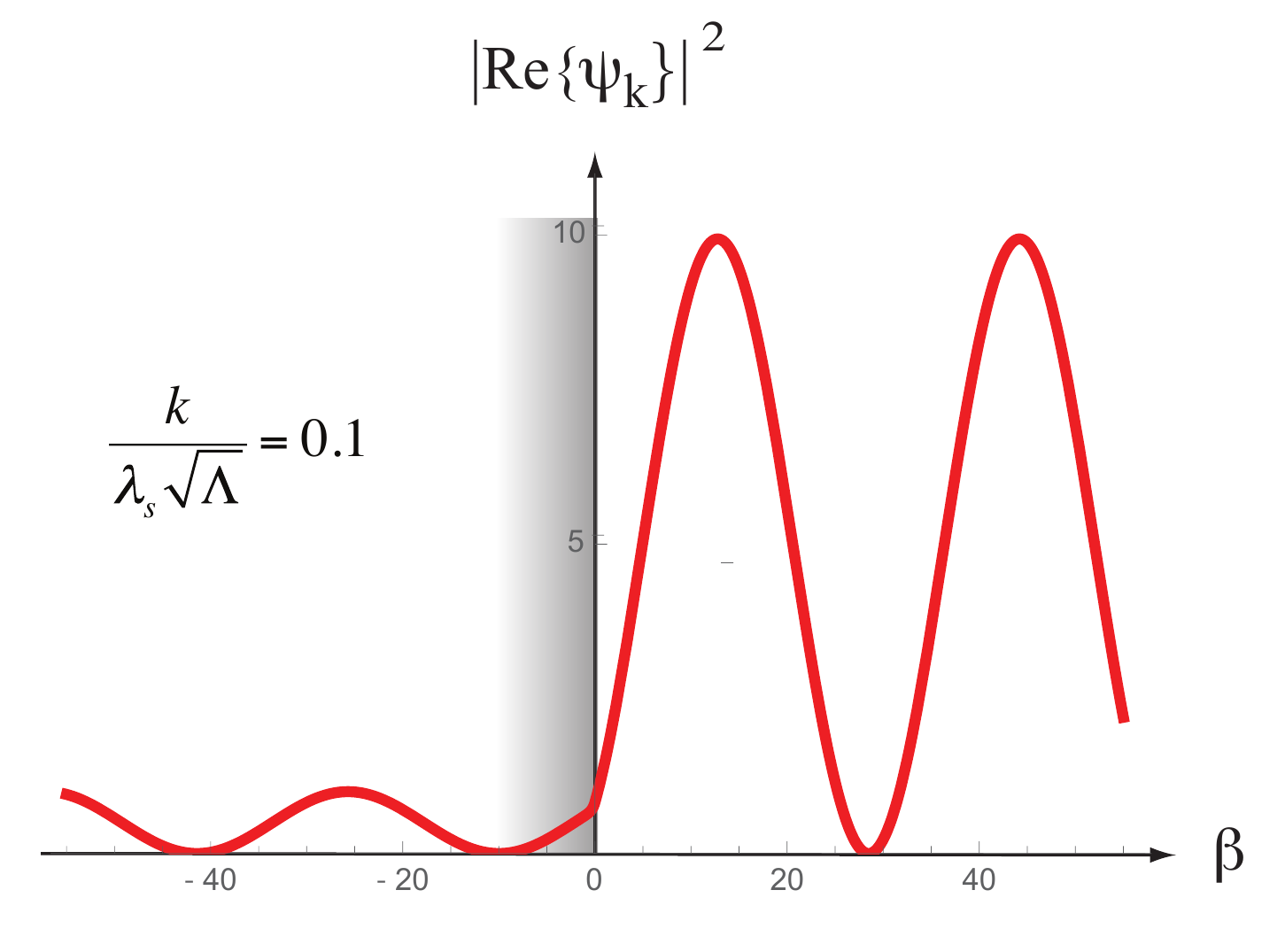}
\caption{Evolution in superspace of the WDW solution which illustrates the anti-tunnelling effect produced by the effective potential barrier (grey shaded region) due to the dilaton potential (\ref{51}). The wave function is not damped but parametrically amplified provided $k<\la_s \sqrt{\La}$, and the effect is larger for smaller $k$.}
\label{f2}
\end{figure}

In order to illustrate this effect we have numerically integrated Eq. (\ref{52}), with the boundary conditions (\ref{53}), for $d=3$ spatial dimensions. The results are shown in Fig. \ref{f2}, where we have plotted the evolution in superspace of the real part of the WDW wave function, for different values of $k$ (the behavior of the imaginary part is qualitatively similar). We have used units where $\la_s^2 \La=1$, so that the effective potential barrier of Eq. (\ref{52}) is non-negligible  only for very small (negative) values of $\b$ (the grey shaded region of Fig. \ref{f2}). Also, we have imposed on all modes the same formal normalization $|\Psi_k|^2=1$ at $\b \ra -\infty$, to emphasize that the amplification is more effective at lower frequency. 

Concerning this last point, we can find an interesting (and reasonable) interpretation of the condition $k<\la_s \sqrt{\La}$ by considering the realistic case of a transition process occurring at the string scale, with $\dot \b \sim \la_s$, with coupling constant $g_s$, and for a spatial region of proper volume $\Om_s$. In that case, by using the result (\ref{49}) for the momentum $k$ expressed in terms of string-scale variables, we can write the condition of efficient parametric amplification in the following form:
\beq k \sim g_s^{-2}(\Om_s /\la_s^d)\,  \laq \,\la_s \sqrt{\La}.
\label{56}
\eeq
It implies that the birth of our present, expanding, post-big bang phase can be efficiently described as a process of anti-tunnelling -- or, in other words, as a forced production of pairs of universes -- from the string perturbative vacuum, in the following cases: initial configurations of small enough volume in string units, and/or large enough coupling $g_s$, and/or large enough cosmological constant in string units. Quite similar conclusions were obtained also in the case discussed in the previous section. 

\begin{figure}[h]
\centering
\includegraphics[width=11 cm]{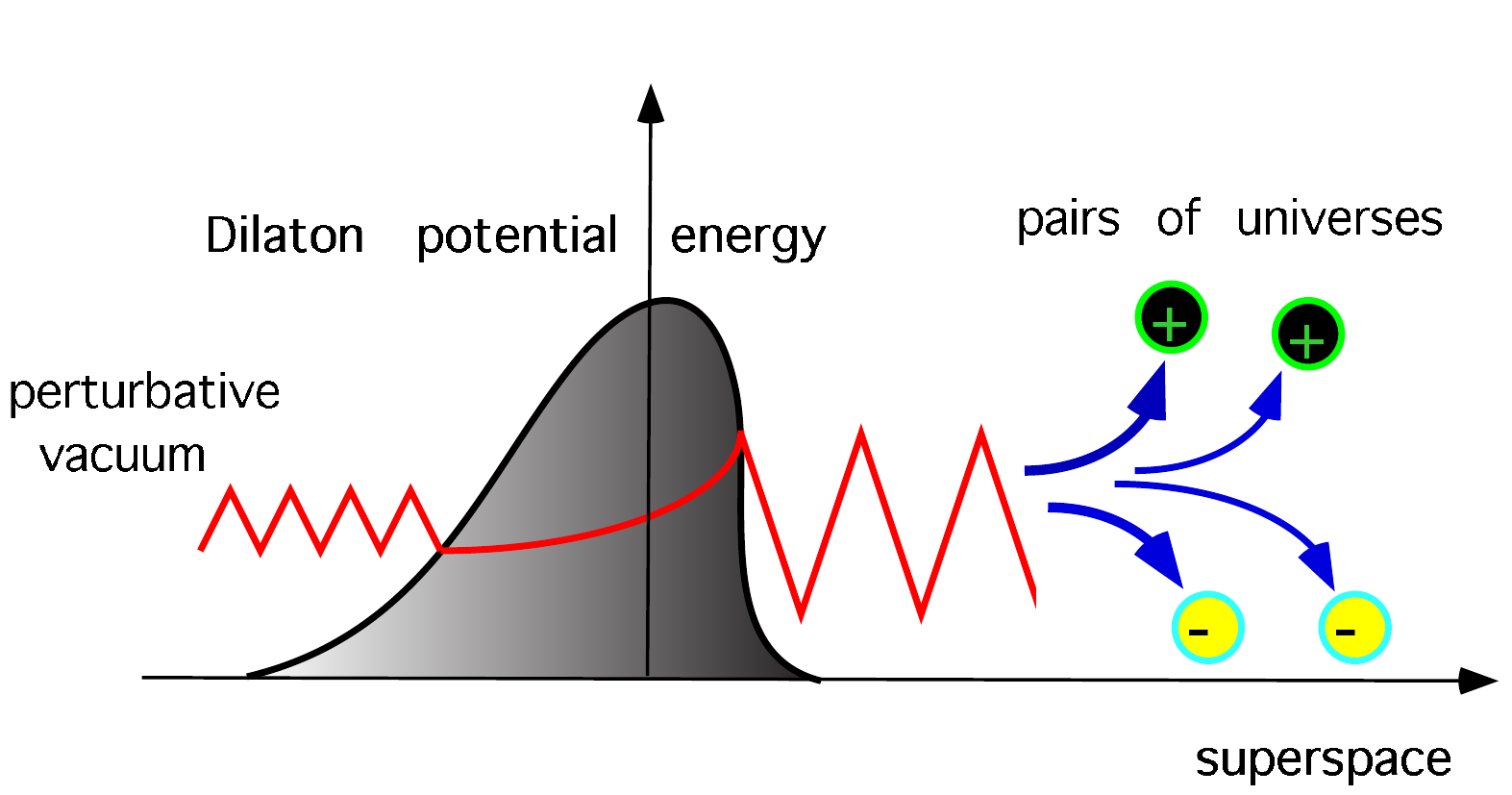}
\caption{Birth of the universe represented as an anti-tunneling (parametric amplification)
effect of the wave function in superspace, or -- in the language of third quantization --  as a process of pair production from the string
perturbative vacuum.}
\label{f3}
\end{figure}

In view of the above results, we may conclude that, for an appropriate initial configuration, and if triggered by the appropriate dilaton potential, the decay of the initial string perturbative vacuum can efficiently proceed via 
parametric amplification of the WDW wave function in superspace, and
can be described  as a forced production of pairs of universes
from the quantum fluctuations. One member of the pair disappears into
the pre-big bang singularity, the other bounces back towards the
low-energy regime. The resulting effect is a net flux of universes that
may escape to infinity in the post-big bang regime (as qualitatively illustrated in Fig. \ref{f3}), with a process which can describe the birth of our Universe as ``anti-tunnelling from the string perturbative vacuum".

\section{Conclusion}
\label{sec6}

The quantum cosmology scenarios reported in this review are based on the low-energy, tree-level string effective action, which is physically appropriate to describe early enough and late enough cosmological phases, approaching, respectively, the initial perturbative vacuum and the present cosmological epoch.

Such an action cannot used to classically describe the high-curvature, strong coupling regime without the
inclusion of higher-order corrections. However, when at least some of
these corrections and of possible non-perturbative effects are
accounted for by an appropriate dilaton potential, the WDW equation
obtained from the low-energy action action permits a quantum
analysis of the background evolution,  and points out new possible
interesting ways for a Universe born from the string vacuum to reach more standard configurations, and evolve towards the present cosmological regime. 
{In such a context, the possible (future) detection of a stochastic background of cosmic gravitons with the typical imprints of the pre-big bang dynamics (see, e.g., \cite{61}) might thus represent also an ``indirect'' indication that some quantum cosmology mechanism has been effective to trigger the transition to the cosmological state in which we are living.} 


\vspace{12pt} 




\acknowledgments{It is a great pleasure to thank all colleagues and friends who  collaborated and made many important contributions to the original research articles reported in this paper. Let me mention, in particular (and in alphabetical order): Alessandra Buonanno, Marco Cavagli\`a, Michele Maggiore, Jnan Maharana,
Carlo Ungarelli, Gabriele Veneziano. This work is supported in part by INFN under the program TAsP ({\it Theoretical Astroparticle Physics}), and by the research grant number 2017W4HA7S ({\it NAT-NET: Neutrino and Astroparticle Theory Network}), under the program PRIN 2017, funded by MUR.}

\reftitle{References}



\end{document}